\documentclass[aps,prb,preprint,superscriptaddress]{revtex4-1}

\usepackage[pdftex]{graphicx}
\usepackage{bm}
\usepackage{color}
\usepackage{textcase, natbib, url}
\usepackage[normalem]{ulem}
\usepackage[colorlinks=true, linkcolor=blue, urlcolor=blue, citecolor=blue]{hyperref}
\usepackage{amsmath,amssymb}
\usepackage{mathtools}

\begin{document}
	\title{Non-universal transmission phase behaviour \\ of a large quantum dot}

	\author{H. Edlbauer$^\dagger$}
	\affiliation{Univ. Grenoble Alpes, CNRS, Grenoble INP, Institut Néel, 38000 Grenoble, France}
	\author{S. Takada$^\dagger$}
	\affiliation{Univ. Grenoble Alpes, CNRS, Grenoble INP, Institut Néel, 38000 Grenoble, France}
	\affiliation{National Institute of Advanced Industrial Science and Technology (AIST), National Metrology Institute of
Japan (NMIJ), Tsukuba, Ibaraki 305-8563, Japan}
	\author{G. Roussely}
	\affiliation{Univ. Grenoble Alpes, CNRS, Grenoble INP, Institut Néel, 38000 Grenoble, France}
	\author{M. Yamamoto}
	\affiliation{Department of Applied Physics, University of Tokyo, Bunkyo-ku, Tokyo, 113-8656, Japan}
	\author{S. Tarucha}
	\affiliation{Department of Applied Physics, University of Tokyo, Bunkyo-ku, Tokyo, 113-8656, Japan}
	\affiliation{RIKEN Center for Emergent  Matter Science (CEMS), 2-1 Hirosawa, Wako-shi, Saitama 31-0198, Japan}
	\author{A. Ludwig}
	\affiliation{Lehrstuhl f\"{u}r Angewandte Festk\"{o}rperphysik, Ruhr-Universit\"{a}t Bochum, Universit\"{a}tsstra\ss e 150, 44780 Bochum, Germany}
	\author{A. D. Wieck}
	\affiliation{Lehrstuhl f\"{u}r Angewandte Festk\"{o}rperphysik, Ruhr-Universit\"{a}t Bochum, Universit\"{a}tsstra\ss e 150, 44780 Bochum, Germany}
	\author{T. Meunier}
	\affiliation{Univ. Grenoble Alpes, CNRS, Grenoble INP, Institut Néel, 38000 Grenoble, France}
	\author{C. B\"auerle $^\star$}
	\affiliation{Univ. Grenoble Alpes, CNRS, Grenoble INP, Institut Néel, 38000 Grenoble, France}
	
	\date{\today}

	\vfill
	\begin{abstract}
			The electron wave function experiences a phase modification at coherent transmission through a quantum dot.
			This transmission phase undergoes a characteristic shift of $\pi$ when scanning through a Coulomb-blockade resonance.
			Between successive resonances either a transmission phase lapse of $\pi$ or a phase plateau is theoretically expected to occur depending on the parity of the corresponding quantum dot states. 
			Despite considerable experimental effort, this transmission phase behaviour has remained elusive for a large quantum dot.
			Here we report on transmission phase measurements across such a large quantum dot hosting hundreds of electrons.
			Using an original electron two-path interferometer to scan the transmission phase along fourteen successive resonances, we observe both phase lapses and plateaus.
			Additionally, we demonstrate that quantum dot deformation alters the sequence of transmission phase lapses and plateaus via parity modifications of the involved quantum dot states.
			Our findings set a milestone towards a comprehensive understanding of the transmission phase of quantum dots.
		
	\end{abstract}
	\vfill
	
		\maketitle

	The phase of the electron wave function lies at the heart of coherent transport phenomena such as universal conductance fluctuations or weak localization\cite{Datta1997,Imry1997,EricAkkermans2011}. 
	One way of accessing this quantity employs a quantum interferometer.
	Interesting phenomena arise when a quantum dot (QD) is inserted into one branch of such an interferometer.
	Then a Coulomb-blockade is present and the wave function of an electron experiences a phase modification at resonant transfer through the QD.
	As one scans through a Coulomb-blockade peak (CP), this transmission phase (TP) gradually changes.
	The magnitude of this TP shift strongly depends on the coupling of the QD to the leads in the interferometer branch.
In the strong coupling regime the TP shift is determined by the electron occupancy of the highest occupied energy level. 
When this level only hosts a single electron, its spin plays a predominant role.
Below a certain temperature threshold, in this case the electron forms a strongly correlated many body state -- the Kondo ground state -- leading respectively to a $\pi / 2$ TP shift across two consecutive CPs\cite{Gerland2000,Takada2014}.
For weak coupling, on the other hand, a TP shift of $\pi$ is theoretically expected to occur along a CP and has been measured for the first time in 1997\cite{Schuster1997}.
The course of this shift can be understood by Friedel's sum rule and follows a Breit-Wigner profile\cite{Yeyati1995}.

	A puzzling situation arises when scanning through several consecutive resonances. 
	In this case, the TP behaviour in between the resonances in principle depends on the spatial 
	symmetries of the QD states\cite{Lee1999,Taniguchi1999}:
	If the involved orbitals have the same parity, a sudden TP lapse of $\pi$ appears in the valley between two consecutive resonances.
	When the orbital parity is changing, on the other hand, such a TP lapse is absent giving rise to a TP plateau.
	Pioneering experiments have investigated the TP across a large QD hosting about 200 electrons\cite{Schuster1997}.
	Surprisingly, for this situation only TP lapses of $\pi$ have been found in between each of the investigated resonances.
	Since the measured series of TP lapses were robust against changes in various QD properties the behaviour was termed \textit{universal}.
	A different behaviour was observed in smaller QDs hosting only a few electrons\cite{Avinun-Kalish2005}.
	With an electron number below ten the occurrence of TP lapses in between the resonances was found to depend on the QD properties.
	Above 14 electrons in the QD, however, only TP lapses were observed giving support for a universal TP behaviour.
	Several theoretical models are devoted to explain the occurrence of TP lapses in a universal regime proposing a mechanism to make the appearance of a lapse more likely for a \textit{larger} QD \cite{Baltin1999, Yeyati2000, Silvestrov2000,Hackenbroich2001,Golosov2006,Bertoni2007a,Karrasch2007,Goldstein2009a,Molina2012,Molina2013,Jalabert2014}.
	Despite these theoretical efforts, there is at present no satisfactory explanation for the \textit{complete} absence of TP plateaus and the question about a universal TP behaviour remains as one of the longest standing puzzles of mesoscopic physics.

	TP measurements are rare due to their experimental difficulty.
	Only a few groups have succeeded in performing such measurements\cite{Schuster1997,Sigrist2004,Aikawa2004,Avinun-Kalish2005} and, therefore, not much data is available to be confronted with theory.
	The recent development of a novel Mach Zehnder type (MZ) interferometer, however, opened a new path for TP measurements\cite{Takada2015,Yamamoto2012}.
	The main advantage of this original design is the suppression of electrons encircling the interferometer structure.
	It avoids multi-path contributions to the phase measurement and ensures reliable two-path interference.
	In this work we employ such a novel electron two-path interferometer to address the question about a universal TP behaviour of large QDs having similar dimensions as in Ref. \citenum{Schuster1997}.
	
	\begin{figure}[!b]
		\includegraphics[width=0.5\textwidth]{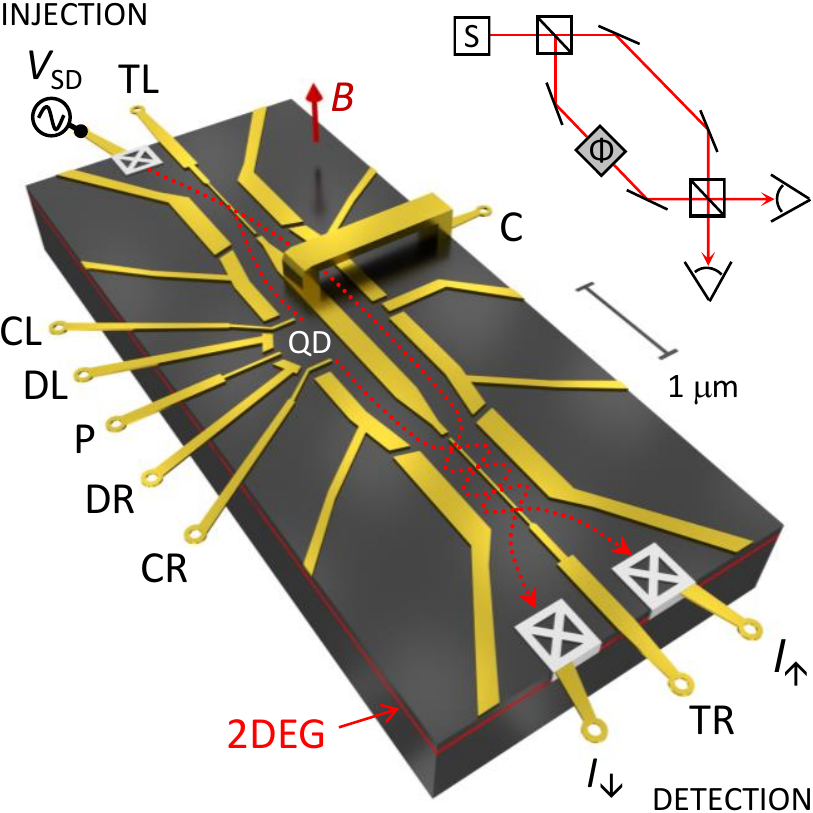}
		\caption{
			{\bf Scheme of the electron two-path interferometer.}
			Shown is a detailed 3D view on the Schottky gates (golden) defining the conductive paths (red, dotted lines) and the interferometer structure in the two-dimensional electron gas (2DEG) located 110 nm below the surface. 
			The ohmic contacts establishing electrical connection to the 2DEG are schematically indicated via the grey crossed boxes in the terminals of the interferometer.
			The inset at the upper right depicts the principle of MZ interferometry -- the photonic counterpart of our electron interferometer.
			\label{fig:setup}
		}
		
	\end{figure}	
	\newpage
	\section*{Results}
	{\bf Experimental set-up.}
	The measurement principle is based on the MZ interferometer.
	Exploiting the Aharonov-Bohm (AB) effect we measure an electronic interference pattern that is used to deduce the TP of the QD.
	Different from the electronic MZ interferometer using edge states in the quantum Hall regime\cite{Ji2003,Roulleau2007,Litvin2007} it is operated at low magnetic fields.
	A scheme of the electron two-path interferometer is shown in Fig. \ref{fig:setup}.
	It is an effective {\it three terminal} device realized in an AB ring, which is sandwiched between two tunnel-coupled wires (TCW):
	The TCW at the injection side (left) serves as a beam splitter guiding injected electrons into the two branches of the interferometer.
	A QD embedded in the lower branch of the AB ring modifies the phase of the electron wave passing through.
	At the detection side (right), a second TCW guides the interfering electron waves -- from the upper and lower interferometer branch -- towards a pair of terminals, where the currents, $I_\uparrow$ and $I_\downarrow$, are measured.
	
\begin{figure}[!h]
		\includegraphics{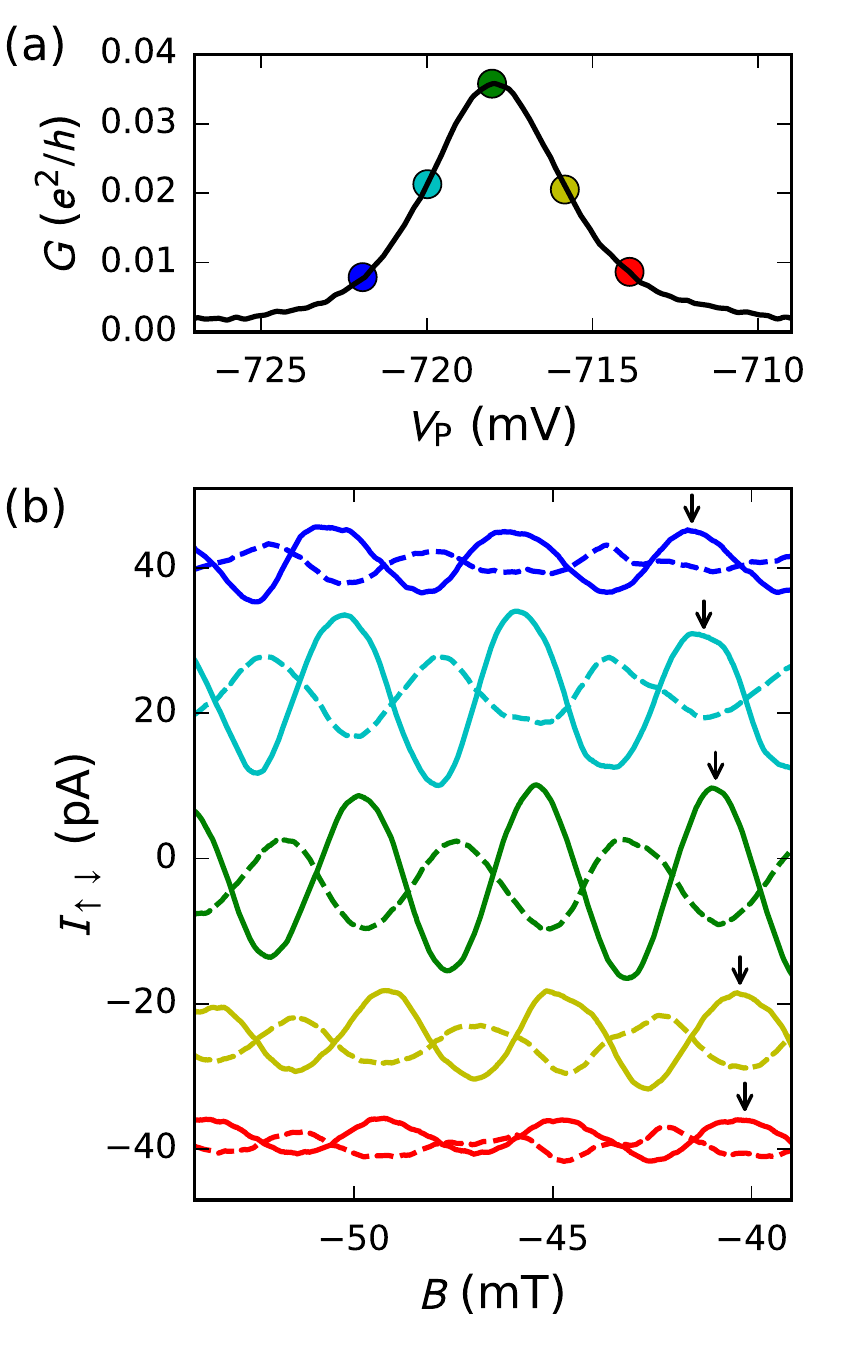}
		\caption{
			{\bf Anti-phase AB oscillations along a Coulomb-blockade peak.}
			(a) Electrical conductance, $G$, as the plunger gate voltage, $V_\textrm{P}$, is swept along the resonance.
			(b) Excerpts of the currents, $I_\uparrow$ (solid line) and $I_\downarrow$ (dashed line), as a function of the magnetic flux density, $B$, for five positions of $V_\textrm{P}$ along the resonance.
			The $V_\textrm{P}$ positions of the current traces are indicated via correspondingly coloured data points in (a).
			The arrows indicate the shift of the AB oscillations.
			For clarity a continuous background is subtracted and an offset is added.  
			\label{fig:analysis}
		}
	\end{figure}

	Changing the flux density, $B$, of a magnetic field perpendicular to the 2DEG, we observe AB oscillations in $I_\uparrow$ and $I_\downarrow$.
	By tuning the tunnel barriers via the voltages, $V_\textrm{TL}$ and $V_\textrm{TR}$, we obtain anti-phase AB oscillations.
	These anti-phase AB oscillations in $I_\uparrow$ and $I_\downarrow$ are the characteristic feature to ensure reliable two-path interference as shown by analytical quantum mechanical calculations, computer simulations and experimental investigations \cite{Yamamoto2012,Bautze2014,Aharony2014,Takada2015}.
	The conductance though the QD is controlled by the so-called plunger gate, P, that is embedded in the QD structure.
	This electrode affects the electrostatic potential of the QD and is used to bring a QD state in resonance with the leads.
	Fig. \ref{fig:analysis}a shows a Coulomb-blockade peak (CP) -- the electrical conductance through the QD as the plunger gate voltage, $V_\textrm{P}$, is swept along a resonance.
	For five positions along the CP the corresponding anti-phase AB oscillations in $I_\uparrow$ and $I_\downarrow$ are shown in Fig. \ref{fig:analysis}b.
	The amplitude of the oscillations is linked with coherent transmission through the QD and, hence, is stronger at the center of resonance.
	As $V_\textrm{P}$ is swept along the resonance, the anti-phase AB oscillations experience a phase shift (see arrows in Fig. \ref{fig:analysis}b), which directly reflects the TP of the QD.

	\newpage
	\newpage
	{\bf Properties of the quantum dot.}
	The QD structure is defined via voltages on six Schottky gates:
	Transversal to the transmission direction the QD is confined by the central electrode, C, and two opposing gates, DL and DR.
	The longitudinal QD confinement and the coupling to the conductive channels of the interferometer branch are defined by two narrow gates, CL and CR.  
	C depletes the two-dimensional electron gas below a central island to form the AB ring.
	The lithographic dimensions of the QD are approximately 0.5 $\mu$m in transverse direction and 0.6 $\mu$m in longitudinal direction. An SEM image of the device is included in the supplementary information.
	From the electron density in the 2DEG and the size of the QD we estimate that the QD contains approximately 300 electrons.

\begin{figure}[!h]
			\includegraphics[scale=1]{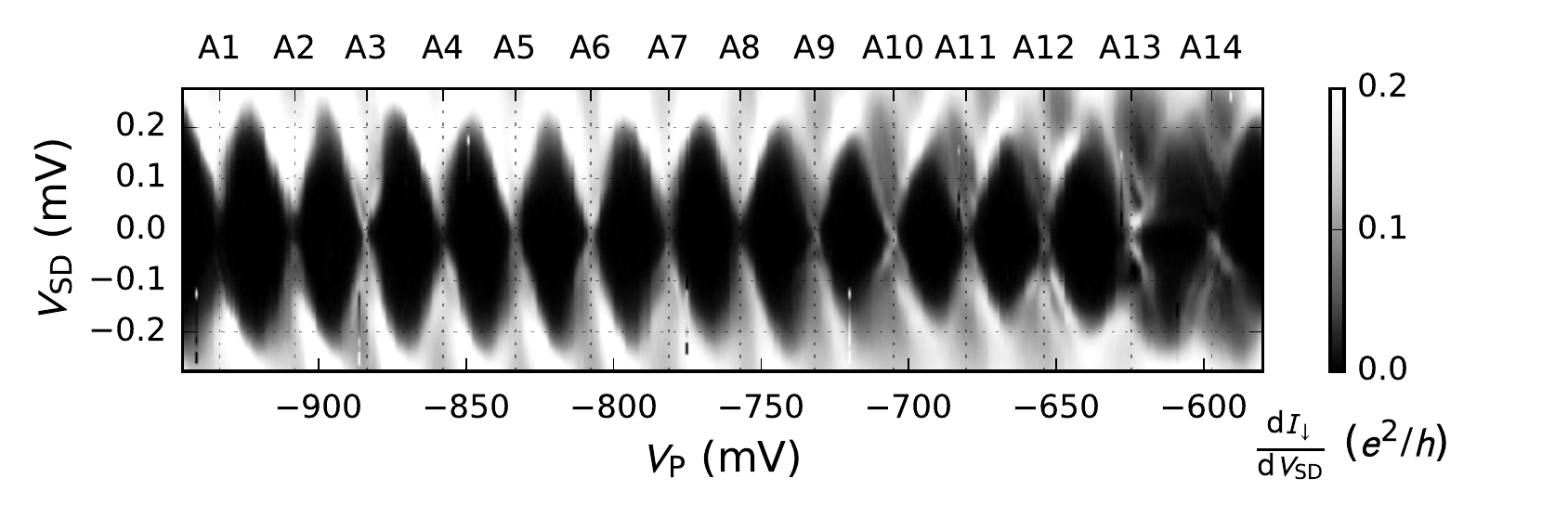}
			\caption{
				{\bf Coulomb diamonds.} Transconductance measurement as the bias voltage, $V_\textrm{SD}$, and the plunger gate voltage, $V_\textrm{P}$, are swept.
				The investigated resonances, A1 to A14, occurring in the swept $V_\textrm{P}$ interval are labeled at the upper axis. 
				\label{fig:diamonds}
			}
		\end{figure}

	Let us now introduce further properties of the investigated QD on the basis of a Coulomb diamond (CD) measurement.
	The presence of the QD causes Coulomb-blockade of electron transport in the lower interferometer branch, which in turn allows to characterise the QD.
	By increasing the tunnel barriers via the voltages $V_\textrm{TL}$ and $V_\textrm{TR}$, electrons are steered only into the lower branch.
	Using lock-in detection, we then measure the transconductance, $dI_{\downarrow}/dV_\textrm{SD}$, to obtain Coulomb diamonds (CD) by stepping $V_\textrm{P}$ and the DC component of the source-drain voltage, $V_{SD}$.
	Fig. \ref{fig:diamonds} shows CD data along 14 resonances (for CPs see also supplementary information).
	We estimate the charging energy, $E_\textrm{C}$, determined by the CD height, as about 240 $\mu$eV. \cite{Hanson2007}
	From the CD width -- the spacing of the resonances --, $V_\textrm{C}\approx 26$ mV, we calculate the voltage to energy conversion factor as $\eta = E_\textrm{C}/V_\textrm{C} \approx 0.01\;e$.
	Approximating the resonances with a Lorentzian function we obtain the coupling energy, $\Gamma \approx \eta \cdot \text{FWHM [V]} \approx 30 \sim 60\;\mu\text{eV}$, from the full width at half maximum (FWHM).
	The energy spacing of the excited states, $\delta$, is hardly resolvable in the measured CD data.
	From the structures appearing between CP A11 and A14, however, we estimate the level spacing as $\delta \approx 60\:\mu\text{eV}$.
	Taking into account our gate geometry and the depth of the 2DEG we can estimate the effective QD area as $A\approx0.3\;\mu \textrm{m}\times 0.4\;\mu \textrm{m}$.
	With this information, we can alternatively estimate $\delta$ from the minimum level spacing of a two-dimensional particle in a box problem\cite{Kouwenhoven1997}. We derive $\delta\gtrsim\hbar^2\pi/mA\approx30\:\mu\text{eV}$, where $\hbar$ is the reduced Planck constant and $m$ is the effective electron mass in a GaAs crystal.
	The two estimates of $\delta$ are quite consistent and	$\delta\approx\Gamma$ implies electron transport at the crossover from the single-level ($\delta>\Gamma$) to the multi-level ($\delta<\Gamma$) regime.
    For the investigated magnetic field range, the Zeeman energy is smaller than temperature fluctuation. We do not observe any enhancement of the valley conductance when lowering the temperature.  Hence, we assume that the Kondo temperature is lower then the electron temperature.
	
	\newpage
	{\bf Transmission phase measurements.}	
	To check the universality of TP lapses in a large QD it is necessary to investigate a set of resonances that is as large as possible.
	When sweeping the plunger gate over a large voltage range, however, crosstalk between the different electrostatic gates defining the QD affects the visibility of the AB oscillations and renders the measurement more difficult.
	To overcome this problem, we split the total sequence of CPs into several measurements.
	For each we carefully fine-tune the voltages on the electrodes defining our interferometer and QD structure to obtain maximal visibility of the anti-phase AB oscillations along the scanned resonances.
	In order to construct the TP along the investigated resonances the data sets have to overlap with at least one CP.
    
\begin{figure}[!h]
    	\includegraphics[width=\textwidth]{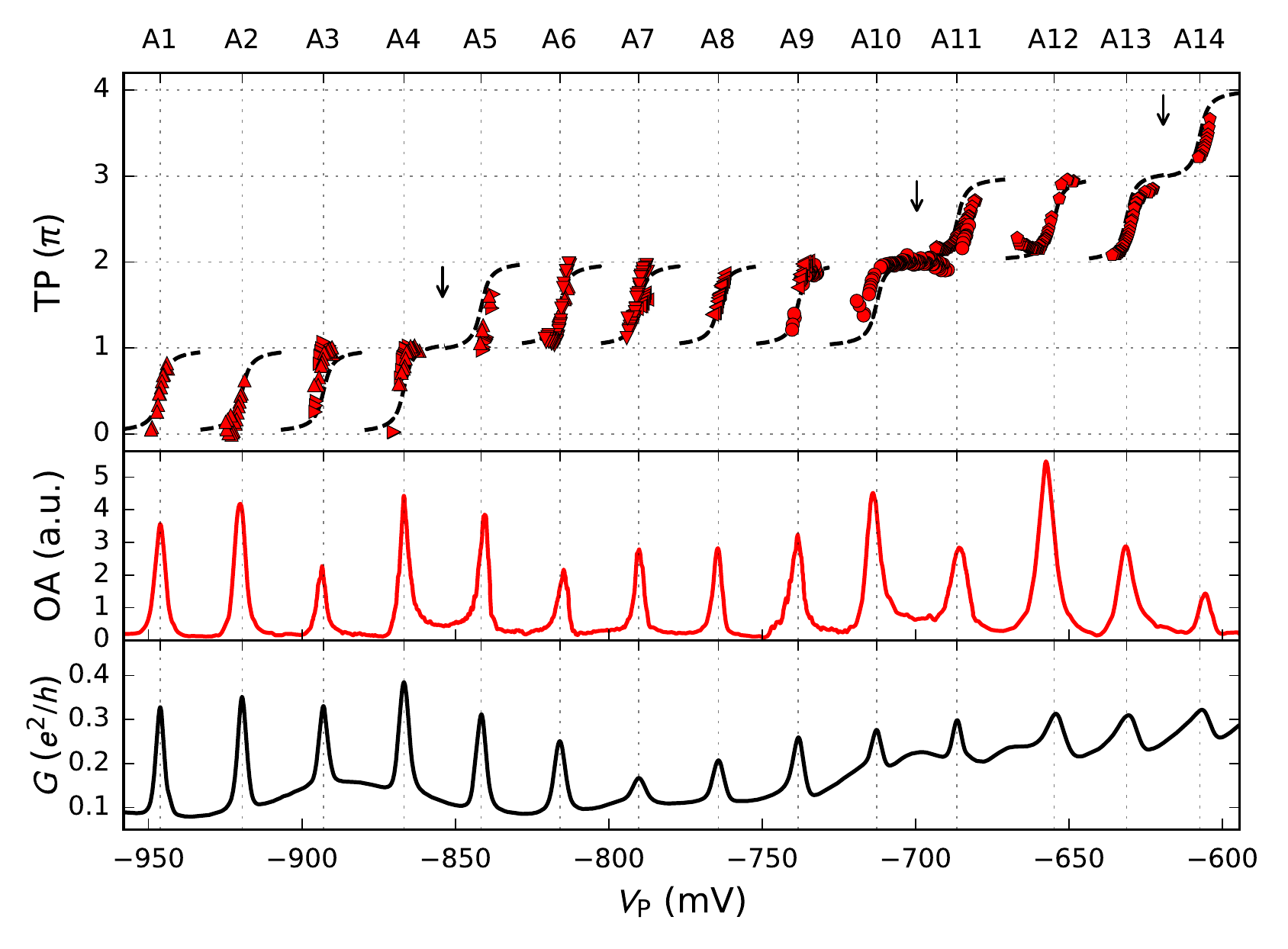}
    	\caption{
    		{\bf Transmission phase along fourteen resonances.}
    		The red data points show the transmission phase (TP) measured along the resonances A1 to A14.
    			The AB oscillation amplitude (OA)  is illustrated in the middle panel (red line).
    			For comparison the electrical conductance, $G=I_\downarrow/V_\textrm{SD}$, is shown in the bottom panel (black line) that is measured at the lower terminal of the interferometer -- the conductance in between the resonances stems from transmission through the upper interferometer branch.
    			The guide to the eye underneath the TP data (dashed line) is constructed from $G$ (see supplementary information).
    		\label{fig:results}
    	}
    \end{figure}

    Following this approach we obtain the TP along fourteen resonances as shown in Fig. \ref{fig:results}.
	Characteristic TP shifts of $\pi$ (red data points) are apparent along all of the investigated resonances labelled as A1 to A14 at the corresponding peaks of conductance, $G$.
    Most importantly, however, the TP data shows a significant signature of {\it non-universal} TP behaviour: among the investigated fourteen resonances three times a TP lapse is absent.
    This leads to clear TP plateaus after the resonances A4, A10 and A13 (see arrows) where the TP cumulates the characteristic shift of $\pi$.
    In between those plateaus, nonetheless, we also observe long sequences of TP lapses as in previous experimental investigations \cite{Schuster1997}.
    
    Theoretically, the occurrence of TP plateaus is linked to a parity change of the QD states\cite{Lee1999,Taniguchi1999}.
For a simple one-dimensional problem, the parity should change when going from one orbital state to the next one. 
For a large QD, which has to be regarded as a two-dimensional system the situation is more complicated. 
In this case the parity is defined by the coupling of the QD state to the two leads. 
One usually defines the   
 quantity  $D_m = \gamma^\textnormal{L}_{m} \cdot \gamma^\textnormal{R}_{m} \cdot \gamma^\textnormal{L}_{m+1} \cdot \gamma^\textnormal{R}_{m+1}$, where $\gamma^\textnormal{L/R}_m$ is respectively the effective coupling at the left and right connection (L/R) of the $m$th QD state\cite{Lee1999,Taniguchi1999,Karrasch2007,Molina2012}.
If two successive QD states, $m$ and $m+1$, have the same parity, this quantity is positive and a singularity point called {\it transmission zero} occurs that causes a TP lapse in between the resonances.
  This leads to a suppression of the AB oscillation amplitude (OA) in the conductance valleys.
   When the parity of two successive QD states changes, on the other hand, the OA in the conductance valley is enhanced.

   This characteristic feature of enhanced OA in conductance valleys without transmission zero is clearly apparent in our experimental data.
   In congruency with a parity-dependent TP behaviour, we find augmented OA in between resonances where a TP plateau occurs (see arrows in Fig. \ref{fig:results}).
   The OA is strongly pronounced in between the resonances A4 and A5 and A10 and A11.
   Our observation of this theoretically expected signature shows that TP measurements indeed can be used to access the sign of the wave functions for the QD states at the connections to the leads.

   Another interesting feature that we find in the present data is the asymmetry of the OA with respect to conductance peaks.
   The asymmetric OA strength around a resonance could indicate spin-flip scattering processes\cite{Akera1993,Akera1999,Koenig2001,Koenig2002}.
By comparing the OA with respect to the conductance peaks, in principle the presence of an empty or partially occupied spin-degenerate level could be deduced\cite{AikawaPRL,Ihn2007}.
   	If the region of reduced OA is located at the positive side of the conductance peak regarding plunger gate voltage, $V_P$, a partially occupied spin-degenerate level is present.
	{\it Vice-versa} is the situation for an empty spin-degenerate level.
   Such a feature is particularly strong at resonances where TP plateaus occur.
   Asymmetric OA peaks are also apparent at several other resonances.
   Nonetheless, we find no systematic correspondence to the occurrence of TP plateaus. 
   Other effects such as the Fano effect\cite{Kobayashi2002,Kobayashi2004,Aikawa2004} or the aforementioned transmission zero can also lead to an asymmetric OA around the resonances.
   The asymmetries of the OA peaks and the sequence of the TP plateaus point out the complexity of the internal structure of the present large QD.
   A detailed characterization of the present sequence of spin-states for the present data is out of reach.
   From the occurence of TP plateaus, however, we can clearly observe changes in orbital parity.

\begin{figure}[!h]
		\includegraphics[width=\textwidth]{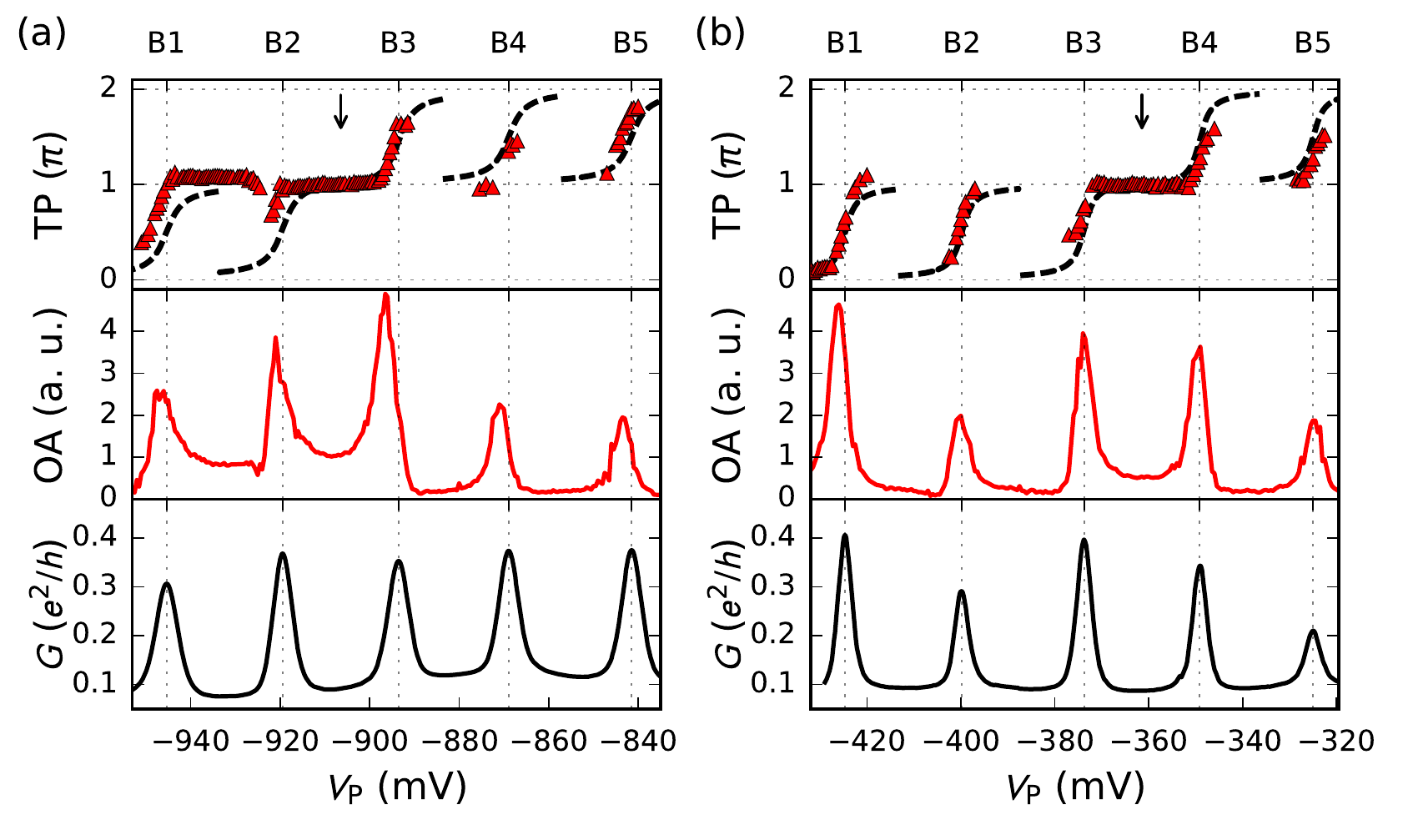}
		\caption{
		{\bf Effect of quantum dot deformation on the transmission phase.}
		Transmission phase (TP) and AB oscillation amplitude (OA) along the same set of resonances (B1 to B5) for different QD configurations:
		(a) Reference with $V_\textrm{DL/R}=-0.92$ V.
		(b) Deformed quantum dot with $V_\textrm{DL/R}=-1.08$ V.
		The deviating intervals of $V_\textrm{P}$ in (a) and (b) result from the different gate voltage applied to the gates $V_\textrm{DL}$ and $V_\textrm{DR}$ and have been chosen such that the same resonances, B1 to B5, are shown.\label{fig:result1}
		The dashed line underneath the TP data serves as guide to the eye and is constructed on the basis of conductance peak data shown in the bottom panel.
		}
	\end{figure}

	{\bf Modifying the quantum dot shape.}
	According to a parity-dependent TP behaviour\cite{Lee1999,Taniguchi1999} one expects modifications of the TP lapse sequence as QD states change. 
	If, for instance, the QD shape is distorted such that the QD states and, thus, the sequence of orbital parities changes, one should observe a modified sequence of TP lapses and plateaus.
	With the present experimental setup we can readily investigate this assertion.
	We deform the QD by changing the balance of the voltages $V_\textrm{DL}$, $V_\textrm{DR}$ and $V_\textrm{P}$.
	As a starting point for the discussion we consider a reference QD configuration with $V_\textrm{DL/R}=-0.92\; \textrm{V}$.
	A set of TP shifts measured along five consecutive CPs is shown in Fig. \ref{fig:result1}a.
	For this reference configuration a TP plateau occurs after resonance B2.
	We modify the QD shape by changing the voltages, $V_\textrm{DL}$ and $V_\textrm{DR}$, to -1.08 V.
	At the same time we adjust the voltage $V_\textrm{P}$ to keep the electron number constant.
	Comparing the TP measurements along the same set of CP for the two situations -- compare Fig. \ref{fig:result1}a and \ref{fig:result1}b -- we find significant changes:
	We observe an alternation of the TP lapse sequence as the QD shape is modified.
	After deformation of the QD the TP plateau now occurs after resonance B3 instead of B2.
	The deformation causes indeed a change in the sequence of orbital parities what is directly reflected in the observed course of the TP.
	According to the altered position of the TP plateau augmented OA appears now in the conductance valley between B3 and B4.
	This change clearly shows the correlation between those two theoretically expected features and indicates a parity change of the QD state that is moved through the bias window at resonance B3.

	 \newpage
	
	\newpage
	\section*{Discussion}

	Several theories have been striving to explain the experimentally claimed universal occurrence of TP lapses\cite{Baltin1999, Yeyati2000, Silvestrov2000,Hackenbroich2001,Golosov2006,Bertoni2007a,Karrasch2007,Goldstein2009a,Molina2012,Molina2013,Jalabert2014}. 
	Nonetheless, a satisfactory explanation for such a universal TP behaviour could not be found. 
	Only for the case where the coupling energy of the QD, $\Gamma$, is larger than the level spacing of the excited states, $\delta$, a universal TP behaviour has been theoretically predicted\cite{Karrasch2007}.
	In experiment, however, such a regime is difficult to achieve.
	So far, all experimental works, including ours, have been done in a regime where $\Gamma \sim \delta $ or $\Gamma < \delta $.
	For these conditions, which are typically encountered in experiment, another set of theory predicts that longer sequences of TP lapses appear for larger QDs, while the  probability of finding a TP plateau remains finite \cite{Molina2012,Jalabert2014}. 

	In line with previous experimental investigations of a QD having similar dimensions we find long sequences of TP lapses.\cite{Schuster1997}
	Scanning the TP along a larger set of successive resonances, however, we also observe the absence of TP lapses giving rise to TP plateaus.
	This finding is in qualitative agreement with the theoretically expected finite probability for TP plateaus of large QDs\cite{Jalabert2014} and clearly shows that the occurrence of TP lapses is not a universal feature as previously claimed.
	The non-universal TP behaviour is underpinned by the demonstration of altered sequences of TP lapses by QD deformation.
	The observed tendency of augmented transmission at TP plateaus gives additional support for TP lapse occurrence depending on the parity of the involved QD states\cite{Lee1999,Taniguchi1999}.
	
	In conclusion, our investigations firmly establish that even for the case of a QD containing a few hundred electrons the absence of TP lapses is possible.
	Our data shows the capability of transmission phase measurements to reveal the microscopic nature of a physical system, such as the orbital parity.
	We anticipate that the present interferometry experiment opens a path for further studies on other fundamental topics, such as correlated electron systems\cite{Thomas1996,Bauer2013,Iqbal2013}.

	\newpage
	\section*{Methods}
	{\bf Experimental setup.}
	Our device is realised with a standard Schottky gate technique in an AlGaAs/GaAs heterostructure hosting a two-dimensional electron gas (2DEG) with a density, $n \approx 3.2\cdot10^{11}\;\text{cm}^{-2}$, and a mobility, $\mu \approx 10^6 \;\text{cm}^2/\text{Vs}$, that is located 110 nm below the surface.
	Applying a set of negative voltages on the electrodes we locally suppress the 2DEG below and define the conductive paths and the interferometer structure.
	Electrical connection to the 2DEG is established via ohmic contacts.
	The central electrode of the AB ring is electrically connected via a bridge electrode deposited on a pad of SU-8 photo-resist.
	The experiments are performed at a temperature of about 30 mK using a $^3$He/$^4$He dilution refrigerator.
	To increase the signal-to-noise-ratio a standard lock-in measurement is performed with a modulation frequency of 23.3 Hz and an amplitude of 20 $\mu$V.
	
	\vspace{5mm}
	{\bf Data analysis.}
	For each value of $V_\textrm{P}$ we obtain the TP by measuring anti-phase AB oscillation of $I_\uparrow$ and $I_\downarrow$.
	The phase data of the AB oscillation is assessed and extracted by the following procedure.
	In order to force the inflection points of the oscillating signal to zero, first, the second derivative of $I_\uparrow$ and $I_\downarrow$ with respect to $B$ is calculated.
	To do so we smooth the data and calculate the first order derivative.
	This operation is similar to subtracting a background signal in order to enhance the quality of the Fourier transform.
	Performing a Fourier Transform of the second derivative of the currents, $I_{\uparrow/\downarrow}^{\prime\prime}$, the AB oscillation amplitude, $\mathrm{OA}_{\uparrow/\downarrow}$, and the phase, $\mathrm{TP}_{\uparrow/\downarrow}$, of the AB oscillation is deduced.
	$\text{OA} = \text{OA}_{\uparrow}+\text{OA}_{\downarrow}$ represents the coherent transmission probability.
	We assess the quality of the TP data by employing two criteria.
	To obtain an acceptable anti-phase oscillation and, thus, mere two-path interference we claim that $|\mathrm{TP}_\uparrow-\mathrm{TP}_\downarrow| = \pi \pm 15 \;\%$.
	Further, the OA has to reach a threshold of 0.5 (a.u.).
	For the data points fulfilling our quality criteria, we then calculate the Fourier transform, $\text{FT}=\mathcal{F}_\textrm{B}(I_{\uparrow}^{\prime\prime}-I_{\downarrow}^{\prime\prime}| B = \Delta B)$ projected on the AB oscillation periodicity, $\Delta B = h / e A \approx 4.5\;\textrm{mT}$, where $h$ describes the Planck constant, $e$ the elementary charge and $A$ the area of the AB ring.
	With this data, we finally extract the transmission phase, $\text{TP} = \arg(\text{FT})$, and the AB oscillation amplitude, $\text{OA} = |\text{FT}|$.

\newpage
\section*{References}
	\bibliographystyle{nature}

\newpage
\section*{Acknowledgments}
We acknowledge fruitful discussion with Rodolfo Jalabert and Dietmar Weinmann and thank Dominique Mailly for teaching us the employed bridge technique.
Sh. T. acknowledges financial support from the European Union's Horizon 2020 research and innovation program under the Marie Sk\l{}odowska-Curie grant agreement No. 654603.
M.Y. acknowledges financial support by PRESTO, JST (No. JPMJPR132D) and Grant-in-Aid for Scientific Research A (No. 26247050). 
S. T. acknowledges financial support by JSPS, Grant-in-Aid for Scientific Research S (No. 26220710), MEXT project for Developing Innovation Systems, and QPEC, the University of Tokyo.
A. L. and A. D. W. acknowledge gratefully support of DFG-TRR160, BMBF - Q.com-H 16KIS0109, and the DFH/UFA CDFA-05-06.
C. B. acknowledges financial support from the French National Agency (ANR) in the frame of its program BLANC FLYELEC Project No. anr-12BS10-001.
This project has received funding from European Union's Horizon 2020 research and innovation program under the Marie Sk\l{}odowska-Curie grant agreement No. 642688.

\section*{Author contribution}

Sh.T., M.Y, S.T, T.M. and C.B initiated this work. H.E. and Sh.T. did the experiment, analysed the data and wrote the manuscript with assistance from T.M. and C.B.. Sh.T. fabricated the sample, G. R. contributed to the experimental set-up. A.L. and A.D.W provided the GaAs heterostructure. All authors discussed the results and commented on the manuscript. C.B. supervised the work.

$^\dagger$ These authors contributed equally to this work

$^\star$ corresponding author: christopher.bauerle@neel.cnrs.fr

\textbf{Supplementary information} accompanies this paper. 

\textbf{Competing financial interests:} The authors declare no competing financial interests.

\newpage

\renewcommand{\thefigure}{{\bf S\arabic{figure}}}
\makeatother
\setcounter{figure}{0}
\setcounter{page}{1}
\thispagestyle{empty}

\begin{center}
\textbf{{\large Supplementary information: Non-universal transmission phase behaviour of a large quantum dot}}\\
\bigskip
\end{center}

\section{Experimental setup}
\begin{figure}[htbp]
		\includegraphics{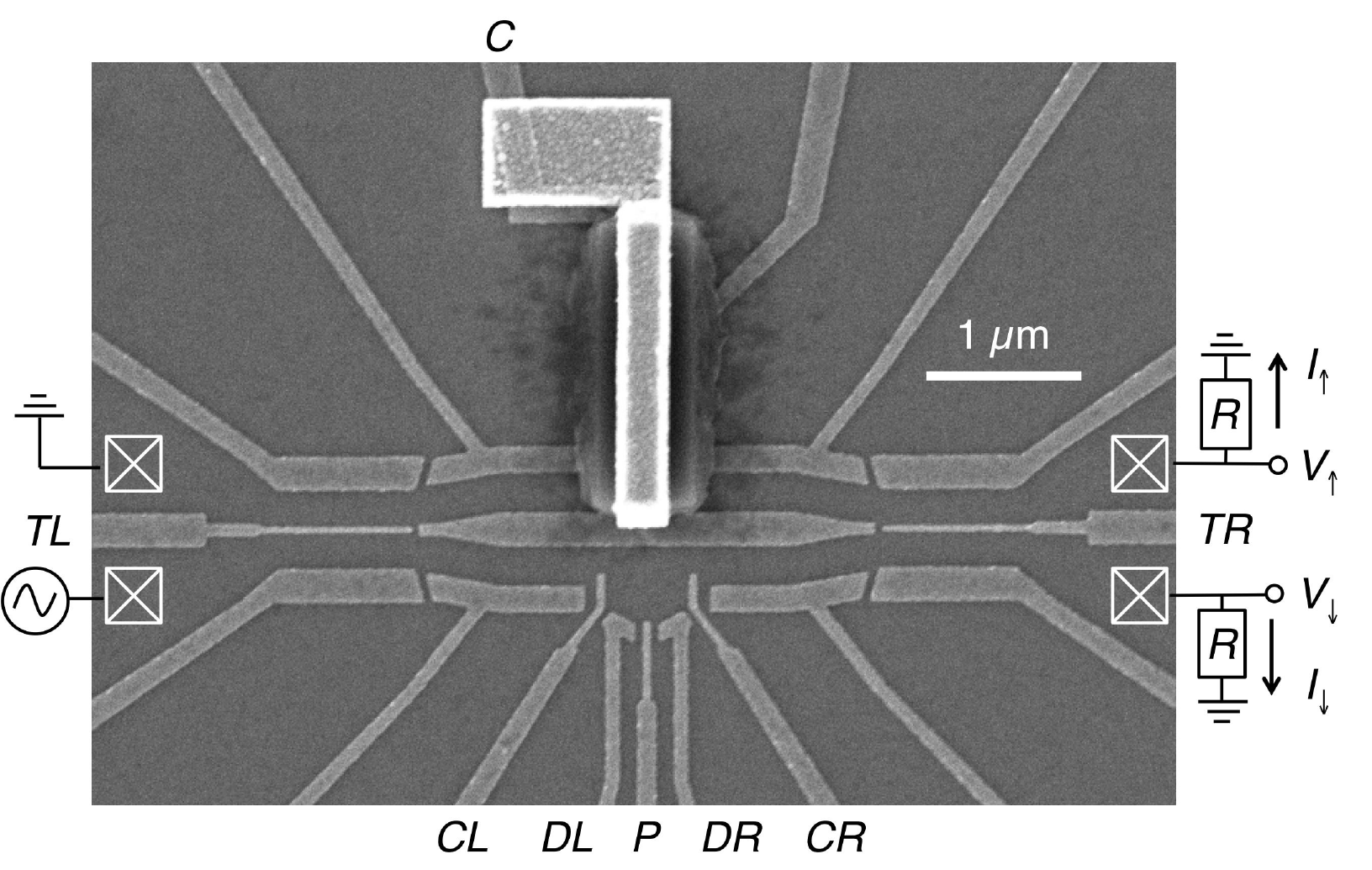}
		\caption{
			{\bf SEM image of the device and scheme of experimental setup.}
			Output currents $I_{\rm \uparrow}$ and $I_{\rm \downarrow}$ are obtained from the voltages across the resistance ($R = V_{\rm \uparrow / \downarrow} / I_{\rm \uparrow / \downarrow} = 10\ {\rm k\Omega}$).
			\label{fig:S1}
		}
	\end{figure}

The employed device is fabricated in a GaAs/AlGaAs heterostructure hosting a two dimensional electron gas (2DEG) 110 nm below the surface.
The structure of the electron two-path interferometer is formed by Schottky gates realized by electron-beam lithography.
A scanning electron microscopy (SEM) image of those electrodes deposited on the surface of the chip is shown in Fig.\,\ref{fig:S1}.
By applying a set of negative voltages on those Schottky gates we locally deplete the 2DEG below and form the potential landscape of the interferometer.
The Aharonov-Bohm (AB) ring is defined by the central electrode, $C$, which is connected via a metal bridge over the upper interferometer branch.
This electrode allows independent control of the tunnel barrier, $TL$ and $TR$.
The metal bridge is fabricated by two additional electron-beam lithography steps: First a pad of SU-8 photoresist is deposited.
This pad prohibits electrical connection of the metal bridge which is deposited in the second step.
A current is injected from the lower left contact by applying an AC voltage (20 ${\rm \mu V}$, 23.3 Hz) and is recovered at the two contacts on the right side. The output currents, $I_{\rm \uparrow}$ and $I_{\rm \downarrow}$, are obtained from the voltages 
$V_{\rm \uparrow}$ and $V_{\rm \downarrow}$ across a resistance of 10 ${\rm k\Omega}$ that is placed on the chip carrier. The experiments are performed in a $^3$He/$^4$He dilution refrigerator at a base temperature of about 30 mK.
\section{Transmission phase measurement and quality criteria}
\begin{figure}[htbp]
		\includegraphics{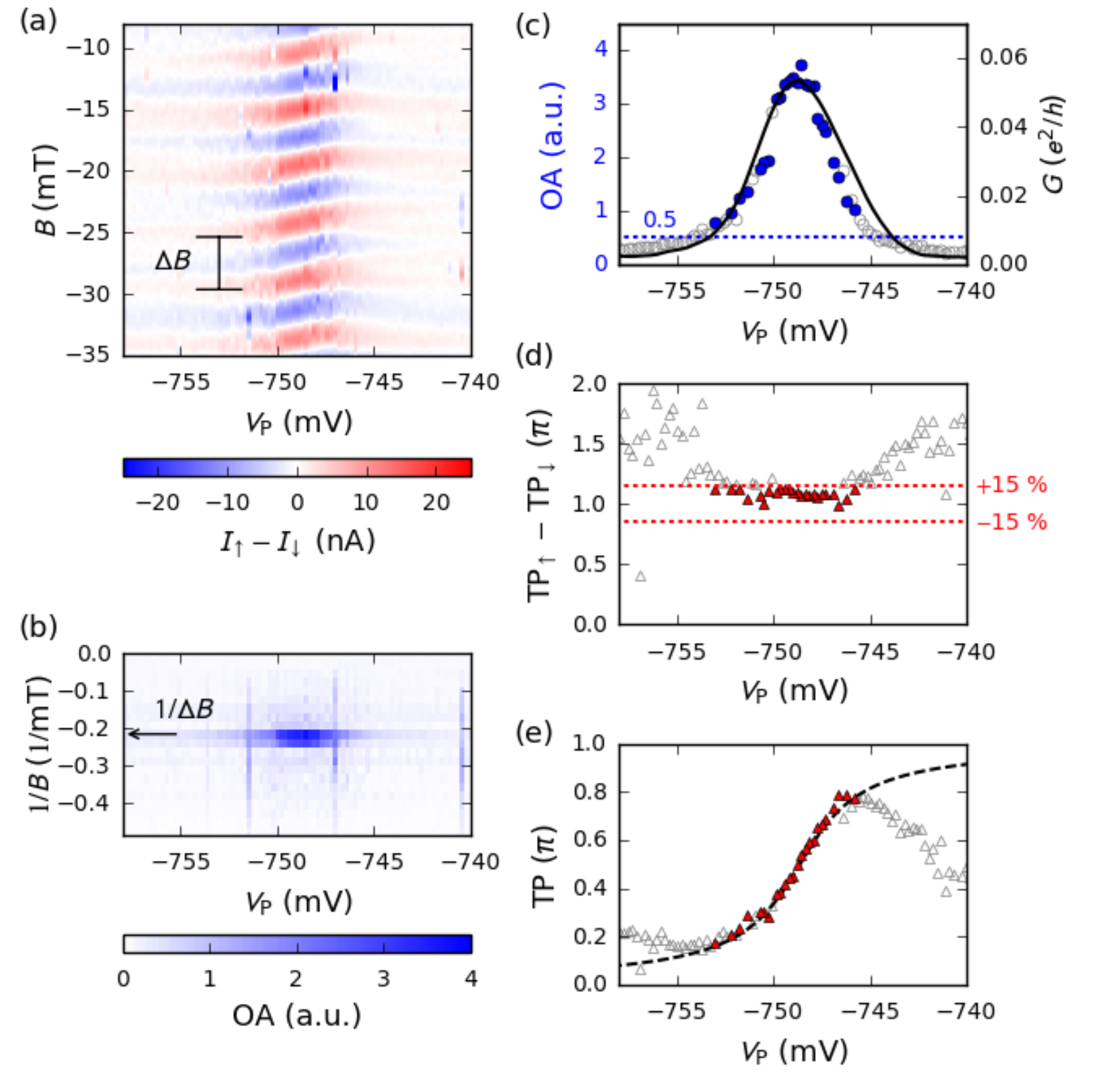}
		\caption{
			{\bf Exemplary set of raw data with analysis.}
			(a) Difference of currents, $I_{\rm \uparrow} - I_{\rm \downarrow}$, as a function of magnetic field, $B$, and the plunger gate voltage, $V_P$. (b) AB oscillation amplitude, OA, obtained by Fourier analysis with respect to $B$. The arrow indicates the periodicity of the AB oscillation, $\Delta B$. (c) Slice of OA at $\Delta B$ (left axis; blue points) and electrical conductance through the QD (right axis; black line) as a function of $V_P$. (d) Difference of the AB phase, ${\rm TP_{\uparrow} - TP_{\downarrow}}$. (e) Transmission phase, TP, (red points) as a function of $V_P$. The dashed  line indicates the TP evolution expected from a Breit-Wigner profile based on a Lorentzian fit of $G$ - see data (c). Data points not fulfilling our quality criteria are shown as open, grey symbols.
			\label{fig:S2}
		}
	\end{figure}

The transmission phase (TP) measurement is performed by sweeping the magnetic field, $B$, and stepping the voltage on the plunger gate, $V_{P}$.
This approach allows to scan the interference fringes via the AB effect while the QD states move through the bias window.
An exemplary set of raw data along a resonance is shown in Fig.\,\ref{fig:S2}a.
Shown is the difference of the currents, $I_{\rm \uparrow} - I_{\rm \downarrow}$, measured at the two terminals at the detection side.
Since $I_{\rm \uparrow}$ and $I_{\rm \downarrow}$ are complementary, the value, $I_{\rm \uparrow} - I_{\rm \downarrow}$, clearly illustrates the AB oscillation.
From the periodicity of the AB oscillation, $\Delta B = h / eA \approx 4.5$ mT, we deduce the effective area of the AB ring as $A \approx 2.14\ {\rm \mu m} \times 0.43\ {\rm \mu m}$, which is consistent with the lithographically defined geometry of the Schottky gates.
The AB oscillations experience a shift as $V_P$ is scanned along a resonance.
This shift directly reflects the course of the TP as a quantum dot (QD) state moves through the resonance.
To obtain the TP data we perform a Fourier transform with respect to $B$.
Since the measured current signals, $I_{\rm \uparrow}$ and $I_{\rm \downarrow}$, contain a continuous background, we smooth the data and calculate the second derivative to force the inflexion points of the AB oscillations to zero:
	\begin{eqnarray}
		I''_{\rm \uparrow / \downarrow} = \frac{\partial^2 I_{\rm \uparrow / \downarrow}}{\partial B^2}.
	\end{eqnarray}
This procedure is similar to the subtraction of a continuous background and leads to similar results.
By forcing the inflexion points to zero, however, the quality of the Fourier transform is strongly enhanced:
	\begin{eqnarray}
		f_{\rm \uparrow / \downarrow} \left(\frac{1}{B},\ V_P\right) = \mathcal{F}_{B} (I''_{\rm \uparrow / \downarrow} (B, V_P)).
	\end{eqnarray}
This approach facilitates the detection of the AB periodicity.
Fig.\,\ref{fig:S2}b shows the oscillation amplitude of the Fourier transformed data, ${\rm OA} = |f_{\rm \uparrow}|+|f_{\rm \downarrow}|$.
At $1 / B = 1 /\Delta B$ the OA data clearly shows a peak corresponding to the AB periodicity, $\Delta B$.
For the further analysis we process only Fourier transformed data projected on $\Delta B$:
	\begin{eqnarray}
		f_{\rm \uparrow / \downarrow}^{\rm AB} (V_P) = \mathcal{F}_{B} (I''_{\rm \uparrow / \downarrow} (\Delta B, V_P)).
	\end{eqnarray}
In order to perform a reliable TP measurement, two criteria have to be met.
The first of those quality criteria ensures sufficient coherent transmission through the QD and, thus, a sufficient signal-to-noise ratio of the AB oscillations.
The second quality criterion assures that multi-path contributions are eliminated via the characteristic feature of anti-phase AB oscillation  in the currents, $I_{\rm \uparrow}$ and $I_{\rm \downarrow}$ \cite{Takada2015}.

The coherent transmission criterion is implemented via a threshold for the AB oscillation amplitude, OA.
Fig.\,\ref{fig:S2}c shows a comparison of a slice of the OA data at the AB peak (blue data points) with the electrical conductance through the QD, $G$ (black line).
$G$ is separately measured by guiding the electrons only through the interferometer branch hosting the QD.
$G$ shows the Coulomb-blockade peak reflecting the total transmission probability through the QD.
The OA data on the other hand is based on the AB effect - a phase coherent quantum interference phenomena.
The OA merely describes the coherent transmission probability and, therefore, serves as adequate quantity for the assessment criterion. In order to ensure sufficient visibility of the AB oscillations we set a coherent transmission criterion such that only data points above a threshold of 0.5 (blue, dashed line) are kept:
	\begin{eqnarray}
		{\rm OA} > 0.5.
	\end{eqnarray}
By this approach only data points with a distinctive AB peak in the Fourier transform are kept.
AB oscillations with a small signal-to-noise ratio, however, are discarded.
The anti-phase criterion assesses the phase difference of the AB oscillation in the currents, $I_{\rm \uparrow}$ and $I_{\rm \downarrow}$.
This AB phase can be directly calculated from the argument of the projected Fourier transform:
	\begin{eqnarray}
		{\rm TP_{\uparrow / \downarrow} = arg}(f_{\rm \uparrow / \downarrow}^{\rm AB}(V_P)).
	\end{eqnarray}
Fig.\,\ref{fig:S2}d shows the phase difference of the AB oscillations in the two terminals for the exemplary data set.
The quality criterion is set such that only data points are accepted where the deviation from the anti-phase relation is below the threshold:
	\begin{eqnarray}
		{\rm |TP_{\uparrow} - TP_{\downarrow}|  < \pi \pm 15\ \%}.
	\end{eqnarray}
Finally, the TP is extracted only for $V_P$ values where the data fulfills the two quality criteria (filled symbols) by evaluating the expression:
	\begin{eqnarray}
		{\rm TP}(V_P) = {\rm arg}(f_{\rm \uparrow}^{\rm AB}(V_P) - f_{\rm \downarrow}^{\rm AB}(V_P)).
	\end{eqnarray}
The TP data for the exemplary data set is shown in Fig.\,\ref{fig:S2}e.
The red, filled triangles show data points fulfilling our quality criteria.
Fitting the conductance, $G$, shown in Fig.\,\ref{fig:S2}c with a Lorentzian function we can construct the expected TP course of Breit-Wigner resonance theory (red line).
The data fulfilling our quality criteria agrees with the expected course, whereas the discarded data points (grey, open triangles) strongly deviate from that course.
\section{Construction of guide to the eye for transmission phase data}
\begin{figure}[htbp]
		\includegraphics{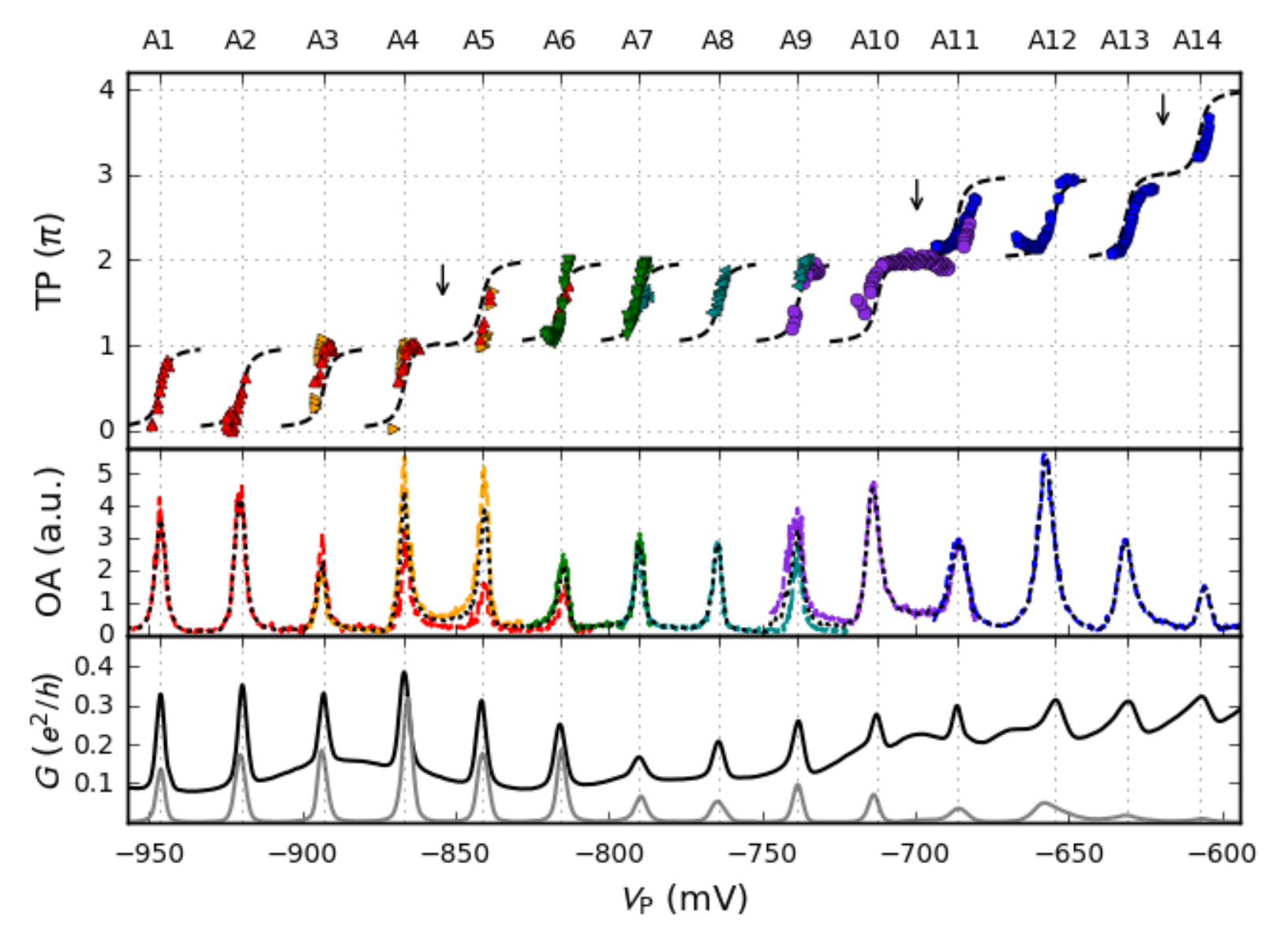}
		\caption{
			{\bf Phase and amplitude of AB oscillations along fourteen resonances.}
			(upper panel) transmission phase, TP, constructed from six measurements along a set of fourteen successive resonances. The individual data sets are indicated via different symbols and colors. The dashed black line in the background is constructed from measurements of conductance, $G$, along the resonances and is used for data alignment and as guide to the eye in the main paper. (central panel) Oscillation amplitude, OA, obtained from Fourier analysis of the individual measurements (colored lines). The dashed line is constructed by convolution of the data sets by a Hann function and is used to represent an approximate FA course. (bottom panel) Conductance, $G$, measured along the investigated resonances at the lower terminal of the interferometer (black line). In addition the bare conductance through the QD is shown, where the electrons are steered only through the lower interferometer branch.
			\label{fig:S3}
		}
	\end{figure}

By fitting the Coulomb peaks with Lorentzian functions one can construct an approximate TP course by applying Breit-Wigner theory.
Then the transmission phase experiences an arctangent shift at each CP depending on the half-width-at-half-maximum, HWHM, and the maximum position, $V_{P,0}$, of the corresponding resonance:
	\begin{eqnarray}
		{\rm TP = arg}\left( \frac{V_P - V_{P,0}}{\rm HWHM} \right).
	\end{eqnarray}
This approach is not profound but provides a reasonable guide to the eye that is presented in the main paper.
The expected TP course is constructed on the basis of conductance data measured at the lower terminal of the interferometer - see black line in bottom panel of Fig.\,\ref{fig:S3}.
Fitting each Coulomb peak with a Lorentzian function, we obtain the positions of the maxima, $V_{P,0}$, and an averaged HWHM value.
Using those fit parameters then we construct an approximately expected TP shift for each resonance - see dashed line in upper plot of Fig.\,\ref{fig:S3}.
Between the resonances TP lapses of $\pi$ are additionally introduced at the positions where the parity of successive QD states does not change in order to have an overlap with the sequence of TP lapses of the measured data.
The characterization of QD is performed by analysing Coulomb diamond data, where the electrons are steered only through the lower interferometer branch.
Measuring the Coulomb-blockade peaks in this condition is not appropriate for comparison with the construction of the guide to the eye for the TP data and for the comparison with the OA due to capacitive crosstalk, which can affect the QD states as the tunnel barriers are closed.
In the bottom panel of Fig.\,\ref{fig:S3} additionally the bare conductance through the QD is shown (grey line), where the electrons are steered only through the lower interferometer branch - as in the Coulomb diamond measurements.
Comparing the bare conductance through the QD (grey line) with the conductance at normal interferometer operation (black line) small deviation of the CP positions are apparent, that stem from the aforementioned crosstalk that slightly affects the QD and occurs as the tunnel barriers are closed - a voltage change of approximately 300 mV.
In the main manuscript, therefore, we use the conductance measured at the lower terminal in normal interferometer mode, where the electrons can pass through both branches.
\section{Data alignment along fourteen successive resonances}
To obtain the TP along a large set of resonances, we first tune the interferometer including the quantum dot to maximise the visibility of the AB oscillations for a maximal number of CPs.
Changing $V_P$, however, affects the coupling of the QD to the two leads in lower interferometer branch.
This crosstalk reduces at some point the visibility of the AB oscillations.
In order to scan through the CP sets of low AB oscillation visibility, we split the measurement and compensate the visibility loss by retuning the interferometer configuration.
To align the fine-tuned data sets regarding phase, it is necessary that the measurements have overlapping $V_P$ intervals.
The data alignment regarding plunger gate voltage, $V_P$, is performed using aforementioned ``guide to the eye" that is based on conductance measurements.
Following this approach with six measurements we obtain the TP course along a set of fourteen resonances.
Fig.\,\ref{fig:S3} shows TP and OA data for each of the conducted measurements - indicated via different colors and symbols.
To obtain the OA course along all of the resonances that is presented in the main paper, we convolute the corresponding OA data sets shown in Fig.\,\ref{fig:S3} with a Hann function - see dashed, black line in lower plot of Fig.\,\ref{fig:S3}.

\section{Magnetic field dependence of Coulomb-blockade peaks}
\begin{figure}[htbp]
		\includegraphics{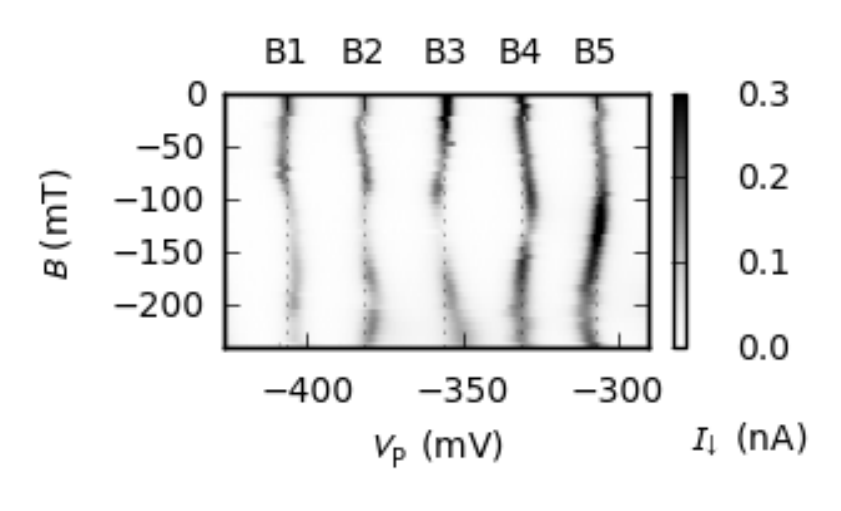}
		\caption{
			{\bf Magnetic field dependence of Coulomb-blockade peaks (CP).}
			Trace of the CP positions in plunger gate voltage, $V_P$, as the magnetic flux density, $B$, is increased from 0 mT to -240 mT. The data is measured at the condition, $V_{DL} = V_{DR} = -1.08$ V.
			\label{fig:S4}
		}
	\end{figure}

As the magnetic flux density, $B$, is swept over a much larger range as in the TP measurements, we observe small shifts of the CP positions in different directions (see Fig.\,\ref{fig:S4}).
The directions of those shifts erratically change as $B$ is increased over several 100 mT.
Since the Land\'{e} $g$ factor for bulk GaAs is -0.44 we expect a minor magnetic field dependence of the CP position on spin.
Rather we expect that the fluctuations of the CP positions are caused by effects of the magnetic field on the orbital wave functions of the QD.
We speculate that strong level mixing occurs and that the erratic changes in the CP shifting direction are caused by orbital energy anti-crossings.
	
\end{document}